\documentclass[apl, pre, twocolumn, amsmath,amssymb]{revtex4-1}
\usepackage{graphicx}
\usepackage{subfigure}
\usepackage{amsfonts}
\usepackage{bm}
\usepackage{dcolumn}
\usepackage{color}
\usepackage{xcolor}

\usepackage{hyperref}

\newcommand{\m }[1]{ \mathbf{#1} }

\newcommand{\nn}{\nonumber }

\newcommand{\rr}{{\mathbf r}}

\newcommand{\w}{\mbox{\boldmath$\omega$}}

\newcommand{\BEQ}{\begin{equation}}
\newcommand{\EEQ}{\end{equation}}
\newcommand{\BEA}{\begin{eqnarray}}
\newcommand{\EEA}{\end{eqnarray}}

\def\beq{\begin{equation}}
\def\eeq{\end{equation}}
\def\beqa{\begin{eqnarray}}
\def\eeqa{\end{eqnarray}}

\newcommand{\mttp}[1]{{\color{black} #1}}

\newcommand{\resp}[1]{\textcolor{black}{#1}}

\begin{document}

\title{
Fractal aggregation of active particles  
}
\date{\today}

\author{M. Paoluzzi}
\email{mttpaoluzzi@gmail.com}
\author{M. Leoni}
\author{M. C. Marchetti }

\affiliation{Physics Department and Syracuse Soft \& Living Matter Program, Syracuse University, Syracuse, NY 13244, USA}

\begin{abstract}
We study active  run-and-tumble particles \resp{in two dimensions} with an additional two-state internal variable characterizing their motile or non-motile state.
Motile particles change  irreversibly  into non-motile ones upon collision with a non-motile particle. The system evolves towards an absorbing state where all particles are non-motile. 
We initialize the system with one non-motile particle in a bath of motile ones and study numerically the kinetics of relaxation to  absorbing state and its structure as function of the density of the initial bath of motile particles and of their tumbling rate. We find a crossover from fractal aggregates at low density to homogeneous ones at high density. The persistence of single-particle dynamics as quantified by the tumbling rate pushes this crossover to higher density and can be used to tune the porosity of the aggregate. At the lowest density the fractal dimension of the aggregate approaches that obtained in single-particle diffusion limited aggregation.  Our results could be exploited for the design of  structures of desired porosity. The model is a first step towards the study of the collective dynamics of active particles that can exchange biological information.
\end{abstract}
\maketitle

\section*{Introduction}
 Self-propelled entities, from active colloids to motile bacteria, show rich collective dynamics 
and emergent patterns ~\cite{Marchetti-Rev-Mod-Phys, Ramaswamy, RevModPhys.88.045006}.
 In these systems transient or permanent
 spatial structures form spontaneously from the interplay of motility and interactions, 
 as for instance in the phenomenon of motility-induced phase separation~\cite{mips},
in collections of  rotors~\cite{Schwarz-Linek,Liebchen}
or in swarms of programmable robots~\cite{Rubenstein}.
 Activity can also be controlled with external perturbations, such as light
~\cite{Stenhammare}, allowing the possibility of controlling active assembly.
Living systems, from bacterial suspensions to tissues, can be thought of as collections of motile active particles that can additionally carry and exchange biological
information or alter their phenotypic or genetic state.  The study of the collective dynamics of such information-carrying active particles has just begun \cite{khadka2018active} and is of great importance for biology \resp{\cite{Heisenberg}.}
 Such information transfer is clearly important for the understanding of   
the  evolutionary behavior of bacterial species~\cite{Keymer} or for regulating intra and intercellular biochemical signaling in multicellular aggregates.
Early work on active gels  has modeled this via the interplay of activity and diffusion
 of molecular species~\cite{Grill}.

\begin{figure*}[t!]
\includegraphics[scale=0.45]{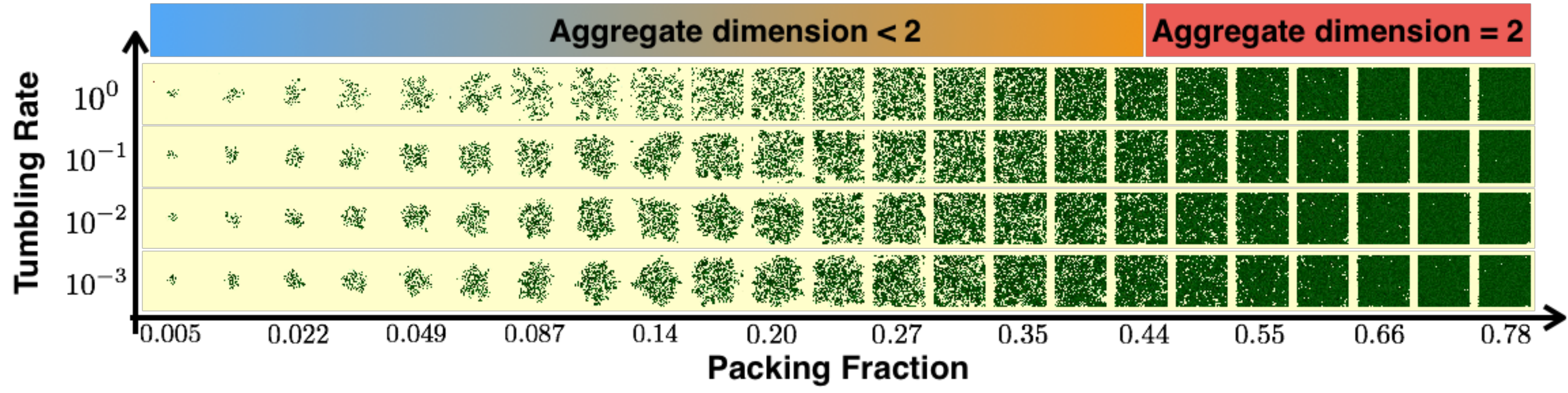} 
\caption{ 
 Snapshots of the absorbing state obtained as the final step of our numerical simulation. 
 The aggregates are homogeneous at high density and porous, with a fractal structure, at lower density. For $\phi$ below about $0.1$ the aggregates no longer fill the box and display the properties of single-particle DLA. 
 }
 \label{fig:structure}
\end{figure*}

In this paper  we  study a minimal model of active
particles with {\it run-and-tumble}  dynamics \resp{in two dimensions} and
an additional,  two-state,
 internal variable $\sigma $ that describes the state of motion: 
 the particles can be motile
 ( $\sigma = 1$) or non-motile or ``dead''  ($\sigma = 0$).
Motile particles change their motility state irreversibly upon collision with non-motile ones.
The system therefore evolves
 towards the absorbing state where $\sigma=0$ for all particles~ \cite{hinrichsen2000non}. 
 The model is relevant to the spreading of infectious diseases
that require person to person contacts ~\cite{Keeling}.
Most previous work in this field has focused on the role of the degree of separation between individuals
and has considered spreading dynamics on a network, with focus on understanding the role of the topological properties of the network, 
especially its connectivity, in controlling disease propagation~\cite{Barrat,Ottino}.  
In the present work, in contrast, we focus on the role of the dynamics of individual agents in controlling the spread of infection. 
\textcolor{black}{Our model may also have some relevance to biofilm formation \cite{Matsushita,Deforet}  and to population dynamics \cite{Deng}.} 
In the latter context the  growing boundary of the $\sigma=0$ state represents
the frontier of an expanding population~ \cite{Murray}.

 The model considered here is a variant of multiparticle diffusion limited aggregation (DLA)~\cite{voss1,voss2}.
 In the classic single-particle DLA process  individual particles are added to the system one at the time and perform a random walk, until they reach the boundary of the aggregate 
 and become part of the cluster \resp{\cite{Witten,Witten2}.} A new particle is added only after the previous one has joined the cluster. In two dimensions ($d=2$) this yields a fractal cluster 
 with fractal dimension $d_f=d-\eta$ and $\eta\simeq 0.3$. The two-point correlation function of the aggregate decays as  a power law $\sim r^{-\eta}$. \resp{Starting with the seminal paper of Witten and Sander \cite{Witten}, several findings about  morphological properties of DLA aggregate have been explored in great detail \cite{meakinDLA1,meakinDLA2,plischkeDLA,barabasi1995fractal}.}
 
 In the \emph{multiparticle} DLA, the aggregate growth takes place in a bath of Brownian particles \resp{with} a small fraction of seed particles \resp{taken to be} initially part of the cluster.
 Multiparticle DLA on a lattice was studied in Refs.~\cite{voss1,voss2} as a function of the concentration of bath particles. It was found that at lo\resp{w} bath concentration the aggregate structure resembled the one
 obtained in single-particle DLA, while a higher concentration led to m\resp{o}re compact clusters.
 
 In this paper we consider an off-lattice 
 model of \resp{multiparticle aggregation where} the bath particles undergo run-and-tumble dynamics, \resp{moving ballistically at speed $v_0$, with the direction of motion randomized at the tumbling rate $\lambda$. By tuning $\lambda$
 the single-particle dynamics 
changes from  diffusive in  the limit
 $\lambda\to \infty$, with $v_0^2/\lambda $ finite \cite{cates_rep}, as implemented in conventional DLA, to ballistic in the limit $\lambda\to0$, as studied in models of ballistic aggregation (BA) \cite{BA1,BA2}.}
\textcolor{black}{Unlike in models of single-particle BA (see, e.g.,  Ref. \cite{kadanoff}), here we do not obtain compact aggregates with fractal dimension equal to the dimensionality of space even  for $\lambda=0$. The reason is that collisions cause the particles' trajectories to deviate from straight runs, generating a finite $\lambda_{eff}(\phi)$. We expect $\lambda_{eff}(\phi)$ to vanish with vanishing density, resulting in compact aggregates for  ballistic single-particle dynamics ($\lambda=0$). This regime is, however, difficult to probe as it requires very long run times and large system sizes.}
 We find that the structure of the final cluster is mainly controlled by the initial concentration of  motile particles. \resp{The tumbling rate promotes the formation of homogeneous aggregates, with an effect similar to that of increasing the random walk step size in  DLA models~\cite{STEP}. It also}  has a profound effect on the time for relaxation to the absorbing state, and increasing $\lambda$ greatly accelerates the dynamics. This result suggest that microorganisms like bacteria may take advantage of run-and-tumble dynamics to speed up aggregation and promote biofilm formation.

\section*{Model }
The model is designed to study the effect of activity on multi-particle aggregation processes \resp{in two spatial dimensions}. 
We consider $N$  \resp{disks} of diameter $a$ in a square box of side $L$ 
with periodic boundary conditions
in two dimensions. All particles interact with short-ranged repulsive interactions.  Initially the system contains a small fraction  of non-motile particles (one particle in the simulations), while all other particles are motile and perform run-and-tumble dynamics. The motile particles switch to a non-motile or ``dead'' state  upon collision with a non-motile particle. At long times the final state will of course be one where all particles are dead. The scope of our work is to quantify the structural properties of the final state and the influence of activity on the kinetics of the aggregation process.
The state of particle $i$, with  $i=1,...,N$, 
is characterized by the position $\rr_i$, a unit vector  $\mathbf{e}_i$
that specifies the  direction or motion during the run phase, and the internal state variable $\sigma_i$, with
 $\sigma_i=0$ for non-motile particles and $\sigma_i=1$ for motile ones.

The dynamics of motile particles is overdamped and governed by the stochastic equation for the translational velocity $\mathbf{v}_i = \dot \rr_i$
and the angular velocity $\w_i = \dot \rr_i \wedge \mathbf{e}_i $, given by \cite{Angelani09,Paoluzzi14,Paoluzzi15}
\BEA \label{micro}
\mathbf{v}_i
&=& v_0 \mathbf{e}_i \sigma_i (1 - \tau_i) + \mu \sum_{i \neq j} \mathbf{f}(r_{ij}),
\\ \nn
\w_i&=& \mathbf{t}^r_i \, \tau_i \, \sigma_i. \,\,
\EEA
In (\ref{micro}), $r_{ij}\equiv | \rr_i - \rr_j|$, $\tau_i$ is an auxiliar state variable, with 
$\tau_i=0$ during the run and $\tau_i=1$ during the tumble, \resp{$\mu=1$ is the mobility.}
During a tumble, particle $i$ acquires a random torque $\mathbf{t}_i^r$ that
rotates the  direction of $\mathbf{e}_i$. Tumbles are Poissonian distributed 
with a rate $\lambda$.
The particles interact mechanically through purely repulsive, short-ranged, forces 
\textcolor{black}{$\mathbf{f}(\rr)=-\nabla_{\mathbf{r}}V(r)$} with 
\resp{ $V(r)=\frac{\epsilon_0 }{12} (\frac{a}{r})^{12}$. In the following we consider $\epsilon_0=1$ }.
 
  The state variable $\sigma_i$
evolves according to
a deterministic rule:
\mttp{when the distance between particle $i$ and particle $j$ at time
$t$ is less than $\delta=\frac{4}{3}a$,}
we update the state of both particles by letting
$\sigma_{i,j}(t+dt)= \sigma_i (t) \cdot \sigma_j(t)$. 

 We have solved numerically Eqs.~(\ref{micro})
using a second-order Runge-Kutta scheme with time step $dt\!=\!10^{-3}$ for
\resp{$v_0 \!=\! 1$}. 
\mttp{The simulations are carried out for 
a maximum of $5 \times 10^{6}$ steps.}
We perform numerical simulations exploring different values of
the tumbling rate $\lambda\!\in\![10^{-3},1]$ and of the
packing fraction $\phi=\rho_0 \pi  (a/2)^2\in[5\cdot 10^{-2} , 0.72]$, with $\rho_0=N/L^2$ the mean number density. The values of $\phi$ are spanned by  varying the number of particles $N=10^2-120^2$ at fixed $L$. 
Most of the
results presented here
have been obtained for systems sizes $L\!=\!120 \,a$.
 We have also examined the dependence of our results on system size considering  $L=180a,240a$.
Finally, we have considered an average over $50$ independent runs.


\begin{figure*}[t!]
\includegraphics[scale=0.4]{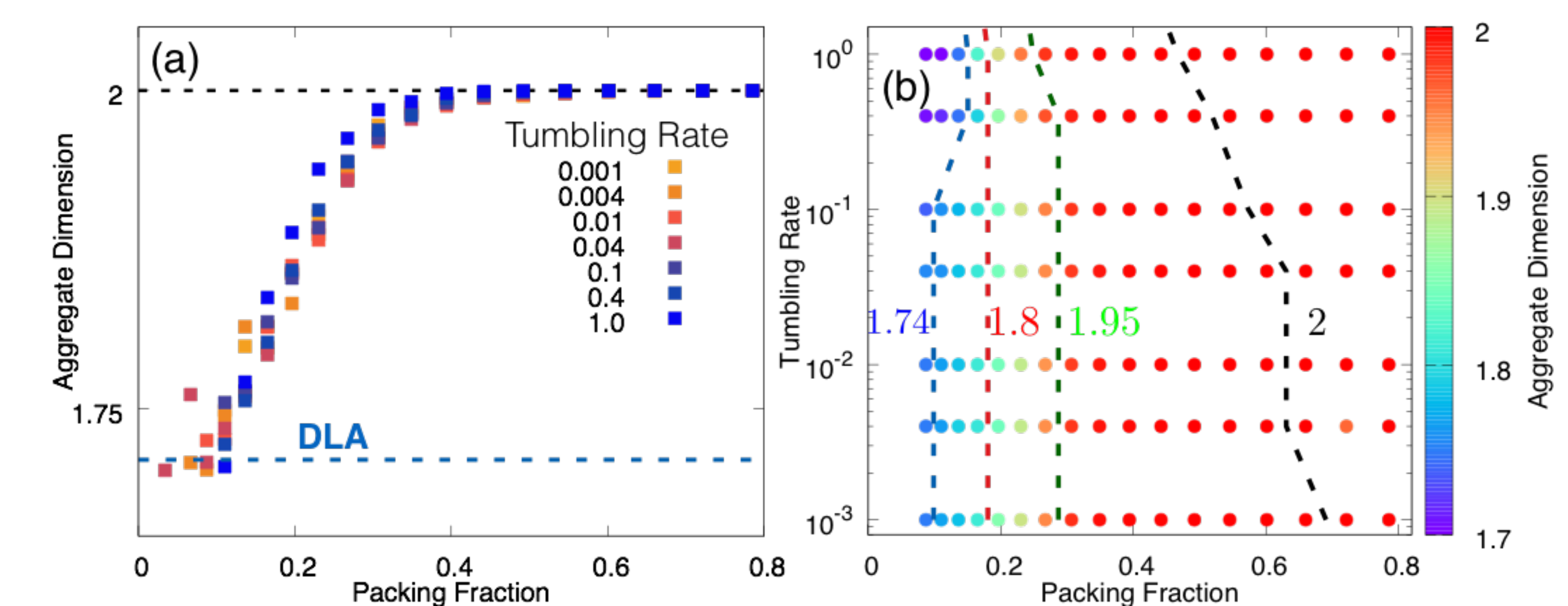} 
\caption{
\textcolor{black}{
(a) 
Fractal dimension $d_f$ of the absorbing state as a function of the packing fraction $\phi$ for tumbling rates 
ranging from $\lambda=0.001$ (orange) to $\lambda=1$ (dark blue). The error bars are smaller than the symbols.
The horizontal blue and black dotted lines correspond to $d_f=1.74$ and $d_f=2$, respectively. 
(b) 
The color shows the fractal exponent $d_f$ of the absorbing state according to the scale indicated identifying various regimes in the $\phi$ versus $\lambda$ plane. The blue and black dotted lines correspond to $d_f=1.74$ and $d_f=2$, respectively. 
\resp{Red and green dotted lines are $d_f=1.8$ and $d_f=1.95$.}
}
}
 \label{fig:fractal-dim}
\end{figure*}

      \begin{figure}[t!]
\includegraphics[scale=0.4]{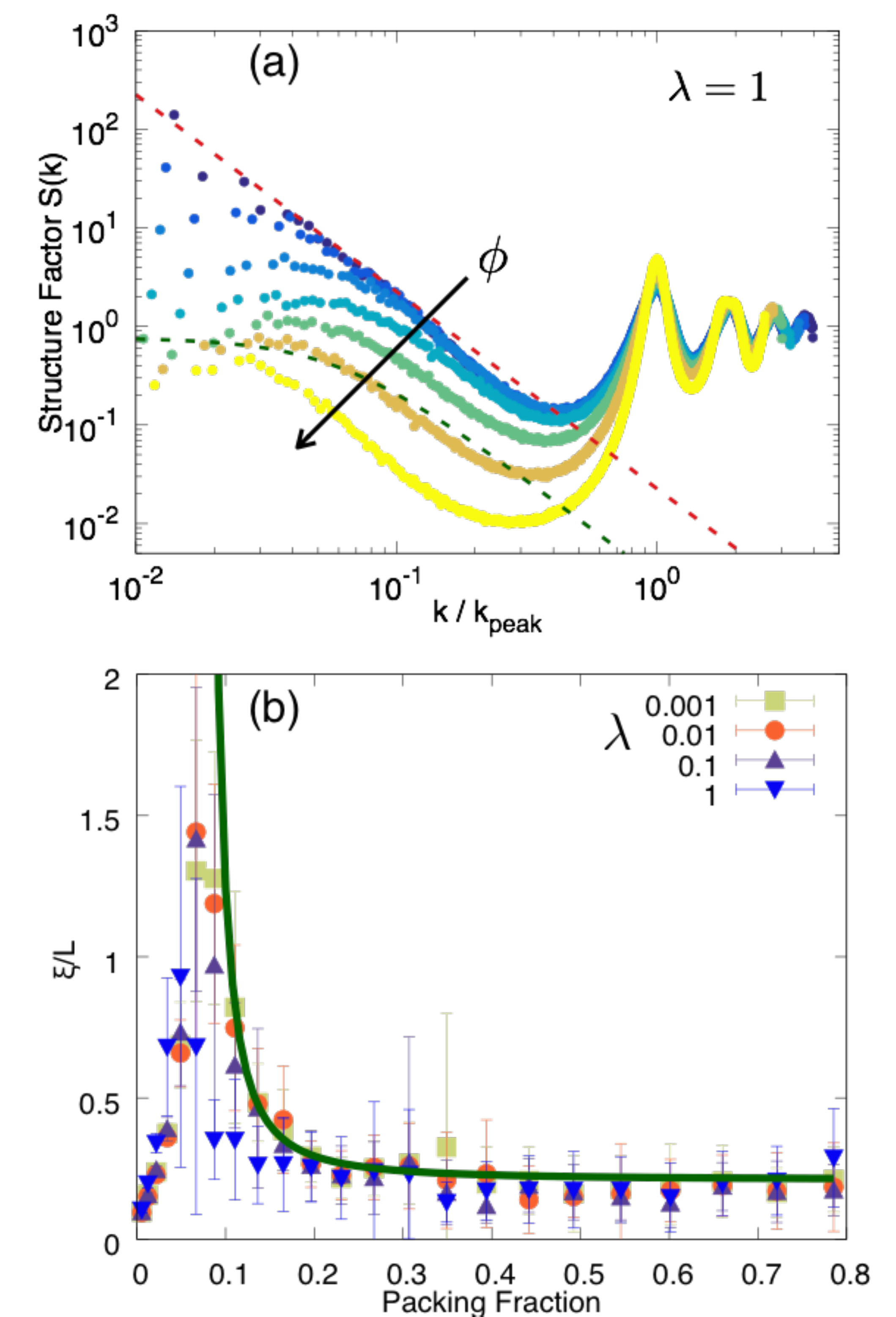} 
\caption{ (a) Static Structure Factor $S(k)$ as a function of $k$ for $\lambda=1$ 
and densities ranging from
$\phi=0.08$ (dark blue) to  $\phi=0.54$ (yellow symbols). 
The black arrow indicates the direction of increasing packing fraction of the initial bath of motile particles. 
The green line is a fit to the Ornstein-Zernike form \mttp{$S(k)=\xi^2 \mttp{S_0} \left[1 + (\xi k)^2 \right]^{-1} $} for $\phi=0.44$, corresponding to $\xi=7.1\pm0.6$. \mttp{This value is in agreement with the estimate of $\xi$ made through Eq. (\ref{corr_xi}) (see panel (b))}. 
The  red line is a fit to a power-law  $S(k)\sim k^{-d_f}$, with $d_f=1.9\pm0.3$ for $\phi=0.087$. 
\resp{Considering the statistical error, the value obtained from the power law is  consistent} with the fractal dimension computed through the box counting method. 
(b) Correlation length $\xi$ in unit of box side $L$ for several $\lambda=0.001, 0.01, 0.1,1$. 
\mttp{The correlation lengths are computed through Eq. (\ref{corr_xi})\resp{, i.e. without any fitting parameter}.}
The green
curve is a fit to $\xi\sim (\phi-\phi_{\text{DLA}})^{-\nu}$, with $\nu=1.7\pm 0.2$. 
\mttp{The data have large error bars. Values of $\xi$ larger than $L$ are due to statistical error.}}
 \label{fig:sq}
\end{figure}

\section*{Results}
All our simulations start with one \mttp{\it{seed}} non-motile particle in the state $\sigma_i\!=\!0$, embedded in a sea of $N-1$ particles with $\sigma_j\!=\!1$ for $i\neq j$.
The final state of course is one where $\sigma_i\!=\!0, \; \forall i$.
To characterize the dynamics we evaluate the time evolution of the fraction of non-motile particles,
${f}(t)=1-N^{-1} \sum_i \sigma_i(t)$, which \mttp{is $1$ when the system reaches the absorbing state, i.e., $\sigma_i=0,$ $\forall i$.}


\subsection{Structure of the absorbing state}
In Fig. \ref{fig:structure} we show snapshots of the final state of non-motile particles at the end of the
 simulations 
 as a function of
packing fraction $\phi$ and  tumbling rate $\lambda$. The seed particle is always at the center of the box.
At high densities the particles fill  the space uniformly \resp{and the spatial dimension of the particle aggregate matches the Euclidean dimension. The value $d_f=2$ is chosen as defining the crossover from fractal to homogeneous structures.} 
At lower density,  the system self-assembles
in porous fractal patterns.
The structure of the final state is quantified by evaluating the fractal dimension of the final non-motile aggregate and the two point correlation function.
To calculate the Minkowski-Bouligand dimension we divide space in a square grid of linear size $\epsilon$. The fractal dimension is 
then defined by evaluating  $\mttp{d_f}(\epsilon) = \ln N(\epsilon)/ \ln \epsilon^{-1}$, where $N(\epsilon)$ is the number
of grid points that are occupied by particles. The fractal dimension $d_f$ is defined as the slope of the linear plot of $\ln N(\epsilon)$ versus $\ln \epsilon$ \mttp{\cite{s1992fractal}}. The results are shown in Fig. \ref{fig:fractal-dim}\mttp{(a)}. 
\textcolor{black}{We have  also verified that  the fractal dimension computed using more general definitions (e.g., information and correlation dimensions~\cite{s1992fractal,barabasi1995fractal}) agrees with that obtained from 
 box counting.
}

The fractal dimension depends only weakly on tumbling rate.
We identify three distinct behaviors as a function of packing fraction. At large packing fraction the final state is homogeneous, with $d_f=2$. At intermediate packing fraction the final aggregate has  porous structure with 
$d_f$ decreasing continuously from $d_f=2$ to the value $d_f=1.74$ (in the limit of large system sizes) that characterizes fractal aggregates obtained in single-particle DLA. At packing fractions below about $0.1$ the aggregates are fractal and compact, in the sense that they do not span the system size. A more quantitative version of the phase diagram in the $(\phi,\lambda)$ plane is shown in Fig. \ref{fig:fractal-dim}\mttp{(b)}. A similar behavior as a function of packing fraction of bath (here motile) particles has been obtained in lattice models of multi-particle DLA ~\cite{voss1,voss2}. In our work the persistence of the dynamics as quantified by the tumbling rate provides an additional knob for tuning the structure of the final state. Smaller tumbling rates, corresponding to more persistent dynamics, enhance the fractal structure of the aggregate by pushing the porous region to higher density.

To further quantify the structural properties of the absorbing state, we have calculated the static structure
factor $S(k)$, defined as~ \cite{hansen1990theory},
\BEQ
S(k)=\frac{1}{N}\left\langle \left| \sum_{j=1}^N e^{-i \m r_j \cdot \m k } \right|^2 \right\rangle\;,
\EEQ
where the angular brackets indicate an average over independent final configurations.
The behavior of $S(k)$ is shown in Fig. \ref{fig:sq}(a) for $\lambda=1$, 
and  packing fractions in the range $\phi=0.08-0.54$.
Since in the absorbing state the particles are always in contact,
$S(k)$ shows a pronounced peak 
at $k_{peak}=2 \pi / a$ even for low initial motile particles density. Counterintuitively, the height of the peak grows with lowering the density, consistent with the growth of the correlation length shown in Fig. ~\ref{fig:sq}(b).
\textcolor{black}{As in colloidal aggregation, see for instance \cite{Gonzalez}, and DLA \cite{Oh,sorensen2001light},  
we expect the structure factor of the final aggregate to obey a scaling ansatz of the form $S(k)\sim\xi^2F(k \xi)$, where $F(x)$ is a universal scaling function, with $F(x\rightarrow 0)={\rm constant}$, and 
$\xi=\xi(\phi,\lambda)$ a correlation length that depends on the model's  parameters. At intermediate wavevectors the structure factor exhibit power-law behavior, $S(k)\sim k^{-d_f}$, as shown in Fig.~3a for one value of packing fraction (see red dashed line). }
%
At high density, where the absorbing state spans the entire system and is uniform, \textcolor{black}{the scaling exponent $d_f$ is replaced by the system's dimensionality $d=2$,} the correlation length is of order $a$ and the structure factor is well approximated by an Ornstein-Zernike form,
$S(k)=\xi^2 \mttp{S_0} \left[1 + (\xi k)^2 \right]^{-1} $\mttp{, with $S_0$ and $\xi$ as fitting parameters,} as shown by the green dashed line in Figs.~\ref{fig:sq}(a) \mttp{for $\phi=0.44$}. 
At lower density, where the structure becomes  porous and fractal, the correlation length grows as the density decreases and one observes the power-law behavior, $S(k)\sim k^{-d_f}$, in a broad range of wave-vectors.  The correlation length extracted from $S(k)$ is shown as a function of packing fraction $\phi$ for various values of tumbling rate in Fig. \ref{fig:sq}(b). 
The correlation length $\xi$ has been calculated by  computing the second-moment of the structure factor according to
~ \cite{amit2005field}
\BEQ\label{corr_xi}
\xi^2=-\left.\frac{\partial \ln S(k)}{\partial k^2}\right|_{k^2=0} \; .
\EEQ
It shows a clear growth with decreasing density, as long as the size $R_g$ of the fractal structure is comparable to the size $L$ of the box. Below packing fractions  $\phi_{\text{DLA}}\sim0.1$, where one obtains DLA-type behavior, $R_g<L$ and $\xi\sim R_g$ decreases with decreasing $\phi$. The growth of $\xi$ with decreasing density for $\phi>\phi_{\text{DLA}}$ can be fit by $\xi\sim (\phi-\phi_{\text{DLA}})^{-\nu}$, where $\nu=1.7\pm 0.2$.
In the porous regime, the correlation length $\xi$ corresponds to the characteristic length scale of the self-similar 
structures, i.e., it quantifies the aggregate's porosity.

  \begin{figure}[h!]
\includegraphics[scale=0.4]{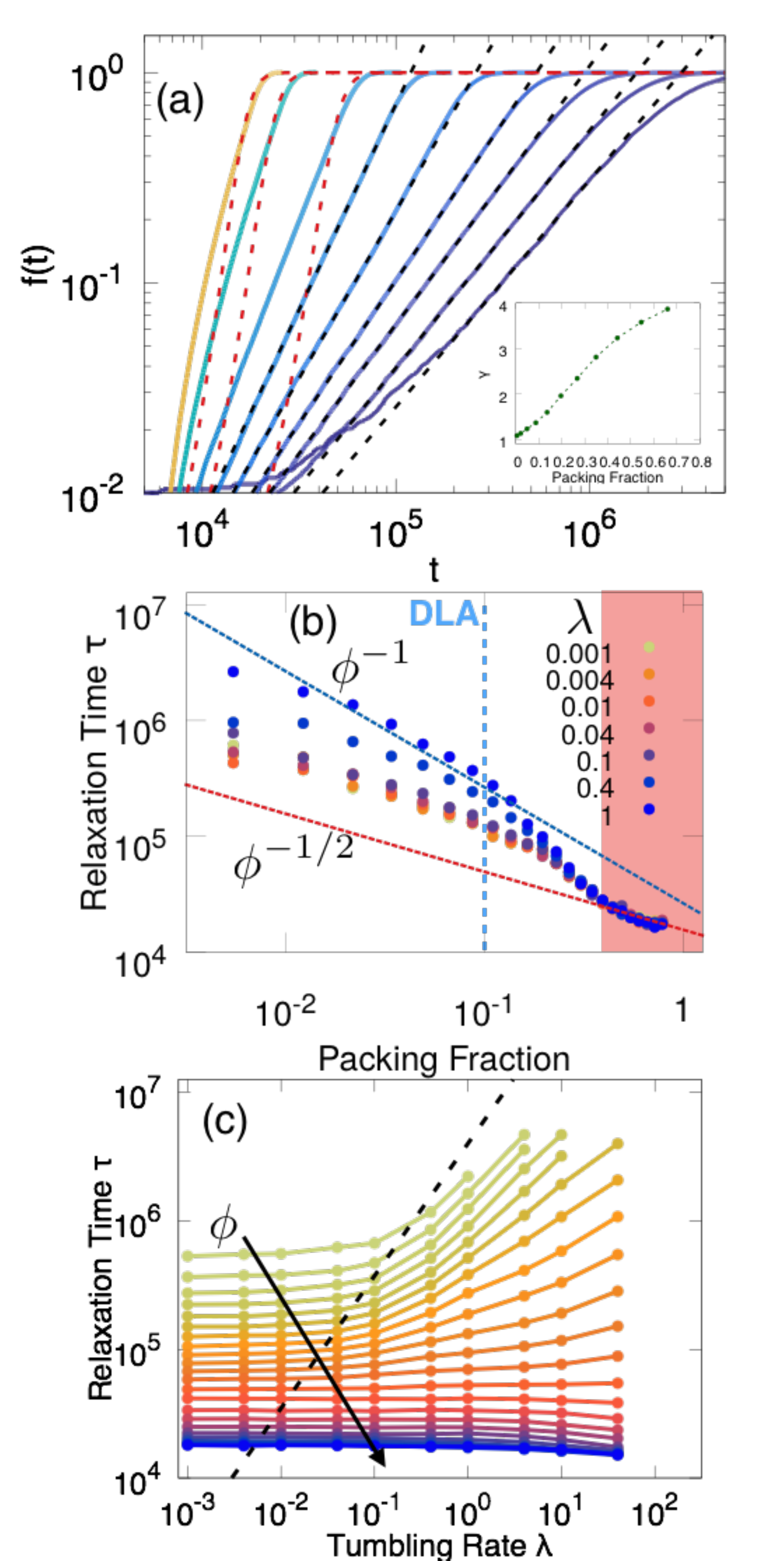} 
\caption{
(a)  Growing of the fraction of non-motile particles in time for increasing packing fraction 
in the porous regime ($\phi=0.05-0.44$) for $\lambda=1$. The red dashed curves are fits to the logistic form given in Eq.~\eqref{eq:m} \mttp{with $\tau$  a fitting parameter}. 
The black dashed lines are fits to $f\sim t^\gamma$. 
Inset: non-universal exponent $\gamma$ as a function of packing fraction $\phi$.
(b) 
Relaxation time $\tau$ defined as $f(\tau)=1$ as a function
of $\phi$ at different values of tumbling rate $\lambda$.
The red and blue dashed lines are fits to the $\phi^{-1/2}$ and $\phi^{-1}$ behavior derived in the text.
The vertical cyan line denotes the value of packing fraction below which the fractal aggregate dimension is $d_{DLA}=1.74$.
(c) Relaxation time $\tau$ as a function of tumbling rate $\lambda$ for a range of packing fractions $\phi=[5 \, \resp{\times}10^{-2},0.72]$. The black dashed line has slope $1$. 
The figure clearly displays the crossover from a regime where the dynamics is ballistic at small $\lambda$ and $\tau$ is independent of $\lambda$ to one where the dynamics is Brownian and $\tau$ grows with $\lambda$. The crossover takes place at increasing values of $\lambda$ as the density increases.} 
 \label{fig:m}
\end{figure}

\subsection{Aggregation kinetics}
The relaxation to the absorbing state is best displayed in terms of the growth of the fraction $f(t)$ of non-motile particles, shown in Fig.~\ref{fig:m}.
At high density the behavior is well described by an exponential growth as obtained from the solution of a logistic equation~\cite{Murray},
 given by
  \begin{equation}
  \label{eq:logistic}
  \partial_t f=\frac{1}{\tau}f(1-f)\;,
  \end{equation}
with $\tau$ a collision time, and solution  
         \begin{equation}
                {f}(t) = \frac{f_0 }{f_0 + (1-f_0 )e^{- t / \tau}  }\;,
         \label{eq:m}
                  \end{equation}
where 
      $ f_0  = {f}(t=0) $ is the initial 
        fraction of non-motile particles (in the numerics, we set $f_0 = 0.01$). 
The logistic model fails, however, at lower density where spatial inhomogeneities are important (see Fig. \resp{\ref{fig:m}(a)} ).
In this case the long time dynamics is best described by a power-law, with $f(t)\sim t^{\gamma(\phi,\lambda)}$. The non-universal exponent $\gamma$ approaches $1$ at low density (see inset in Fig. \ref{fig:m}(a)).
The main effect of activity, as compared to the Brownian dynamics considered in Ref.~\cite{voss1,voss2}, 
is to accelerate the approach to the absorbing state (see Fig. \ref{fig:m}(c)).

We show in Fig.~\ref{fig:m}(b,c) the relaxation time  $\tau$ defined as the time needed for the system
to reach the final configuration where $\sigma_i=0$, $\forall i$, i.e., $f(\tau)=1$.
The relaxation time decreases with increasing density and becomes independent of  $\lambda$ at high density. This crossover is most evident in Fig.~\ref{fig:m}(c) that shows $\tau$ as a function of tumbling rate for a range of densities.
This behavior can be understood qualitatively as a crossover from persistent (or active) to Brownian dynamics using the following simple argument.
The mean-square displacement ${\rm MSD}(t)=\langle[\vec{r}(t)-\vec{r}(0)]^2\rangle$ of an individual particle performing run-and-tumble dynamics displays a crossover 
from ballistic behavior ${\rm MSD}(t)=v_0^2t^2$ for $t<\lambda^{-1}$ to diffusive behavior ${\rm MSD}(t)=4Dt$ for $t<\lambda^{-1}$, with  $D=v_0^2/(2\lambda)$ an effective diffusivity~\cite{MCM-colloid}. The dynamics can be characterized by an associated 
 persistence length $\ell_p=v_0/\lambda$ over which individual particles travel ballistically in a straight line. The time scale $\tau$ controlling collisions that transform motile particles into non-motile ones can then be estimated by ${\rm MSD}(\tau)=\phi^{-1}$, where $\phi^{-1/2}$ is the mean separation between particles. When $\ell_p>\phi^{-1/2}$, the dynamics is ballistic and this gives $v_0^2\tau^2\sim\phi^{-1}$, or $\tau\sim\phi^{-1/2}$ independent of $\lambda$, as shown in Fig.~\ref{fig:m}(c).
 When $\ell_p>\phi^{-1/2}$ the dynamics is diffusive, with the result $D\tau\sim\phi^{-1}$ or $\tau\sim\lambda/\phi$.
 The results shown in Fig.~\ref{fig:m}(b,c) are qualitatively consistent with this simple argument.

 \section*{Concluding remarks}
We have examined the collective dynamics of repulsive 
 run-and-tumble particles that exchange motility state upon contact 
and evolve
 irreversibly towards an absorbing steady state of
non-motile particles. The structure of the absorbing state changes from fractal to homogeneous as the density of the initial bath of motile particles increases - a behavior qualitatively similar to the one previously observed in models of multi-particle DLA~\cite{voss1,voss2}. 
The aggregates evolve from  fractals at low density and high tumbling rates,
 as in diffusion limited aggregation~\cite{Witten}, to homogeneous structures, as typical of clusters in Eden's models~\cite{eden1961}. 
 The persistence of the dynamics controlled by the tumbling rate provides a new knob for tuning the structure of the aggregate, with persistence accelerating the relaxation and promoting uniform structures.
  
By examining both the structure of the absorbing state and the relaxation dynamics when varying the packing fraction, $\phi$, of the initial bath of run-and-tumble particles and their tumbling rate, $\lambda$, we identify three regimes:
\begin{enumerate}
\item[(i)] At low density ($\phi <\phi_{DLA}\sim 0.1$) the absorbing state is a fractal with dimension $d_f\simeq 1.74$ as in single-particle DLA. In this regime the correlation length is bound by the size $R_g<L$ of the cluster: it grows with density and it approaches the system size $L$ as $\phi$ approaches $\phi_{DLA}$ from below. The persistence of the single particle dynamics renders $\phi_{DLA}$ a weak function of $\lambda$, pushing it to lower density with increasing persistence.
\item[(ii)] At intermediate density the aggregate is a space-filling porous structure of fractal dimension growing smoothly from $d_f=1.74$ to $d_f=2$ with increasing density. The relaxation to the absorbing state is power-law, with a non-universal exponent. Persistence of the single-particle dynamics accelerates the relaxation and promotes the formation of homogeneous aggregates.
\item[(iii)] At high density the final aggregate is homogeneous with fractal dimension $d_f=2$. In this regime the relaxation to the absorbing state is exponential and well described by a simple logistic model.
\end{enumerate}

Our model provides a first step towards the study of the effects of motility on population dynamics \mttp{\cite{lavrentovich2013radial,lavrentovich2016spatially,PhysRevLett.119.188003}.} 
In this context it will be interesting to allow for processes where motile particles can be reactivated or `reawakened' after a lag time. 
Our preliminary work on such more general models suggest a rich behavior with the possibility of survival of a population of motile particles in the steady state.
It may be possible to test  our predictions on the role of persistence of the single-partcile dynamics in tuning the structure of the steady state aggregates by
employing light to locally affect the tumbling rate of
genetically modified swimming bacteria \cite{poon,frangipane18}.
The mechanism described here could also be exploited  to use active colloids, where persistence is tuned by rotational diffusion, to design and 
assemble  micro-structures of desired porosity
 for technological and biomedical applications.

\section*{Acknowledgments}  
This work was supported by the Simons Foundation Targeted Grant in the Mathematical Modeling of Living Systems Number 342354 (MP, MCM) and by the National Science Foundation through award DMR-1609208 (MCM). All authors thank the Syracuse Soft \& Living Matter Program for additional support and David Yllanes for illuminating discussions.

 \bibliography{biblio-info-active}

\end{document}